# Gas dependent hysteresis in MoS$_2$ field effect transistors

**Francesca Urban**[1,2], **Filippo Giubileo**[2], **Alessandro Grillo**[1], **Laura Iemmo**[1,2], **Giuseppe Luongo**[1,2], **Maurizio Passacantando**[3], **Tobias Foller**[4], **Lukas Madauß**[4], **Erik Pollmann**[4], **Martin Paul Geller**[4], **Dennis Oing**[4], **Marika Schleberger**[4,*] **and Antonio Di Bartolomeo**[1,2,5,*]

[1]Physics Department "E. R. Caianiello", University of Salerno, via Giovanni Paolo II n. 132, Fisciano 84084, Italy

[2]CNR-SPIN Salerno, via Giovanni Paolo II n. 132, Fisciano 84084, Italy

[3]Department of Physical and Chemical Science, University of L'Aquila, and CNR-SPIN L'Aquila, via Vetoio, Coppito 67100, L'Aquila, Italy

[4]Faculty of Physics and CENIDE, University of Duisburg-Essen, Lotharstraße 1, Duisburg 47057, Germany

[5]Interdepartmental Centre NanoMates, University of Salerno, via Giovanni Paolo II n. 132, Fisciano 84084, Italy

E-mail: adibartolomeo@unisa.it, marika.schleberger@uni-due.de

**Abstract**

We study the effect of electric stress, gas pressure and gas type on the hysteresis in the transfer characteristics of monolayer molybdenum disulfide (MoS$_2$) field effect transistors. The presence of defects and point vacancies in the MoS$_2$ crystal structure facilitates the adsorption of oxygen, nitrogen, hydrogen or methane, which strongly affect the transistor electrical characteristics. Although the gas adsorption does not modify the conduction type, we demonstrate a correlation between hysteresis width and adsorption energy onto the MoS$_2$ surface. We show that hysteresis is controllable by pressure and/or gas type. Hysteresis features two well-separated current levels, especially when gases are stably adsorbed on the channel, which can be exploited in memory devices.

**Introduction:**

During the last few years, there has been an increasing interest in two-dimensional (2D) materials for technological applications. The presence of a tunable and layer-sizable bandgap, the mechanical strength and the chemical and thermal stability make 2D transition metal dichalcogenides (TMDs) good candidates for next-generation electronic devices [1–6]. Theoretical and experimental studies have demonstrated that 2D TMDs based devices can achieve carrier mobilities up to hundreds $cm^2V^{-1}s^{-1}$, very high on/off ratios up to $10^8$, low power consumption and short switching times [7–14]. In their 2D form, owing to the low density of states, TMDs enable enhanced gate control in easy-to-fabricate transistors immune from short-channel effects. Indeed, compared to silicon traditional devices, TMD field-effect transistors (FET) show steeper subthreshold swing (SS), negligible drain-induced barrier lowering (DIBL), high drive current capabilities and low standby off-current, even when relatively thick gate oxides are used [15,16]. The control of *n*- or *p*-type conduction is another important advantage offered by TMDs transistors. Indeed, ambipolar conduction and high on/off ratio are important features for stable low-power consumption and performant logic applications [17,18]. Furthermore, TMDs can be integrated into silicon fabrication technologies to realize devices with nanometric channel length, suitable for high-density integrated circuits.

Additional features like photoconduction [6,19] and spin-orbit splitting [20] have been investigated for optoelectronic and spintronic applications [21]. Moreover, the sharp edges of TMD flakes, combined with intrinsic doping and low electron affinities, make TMDs promising materials also for field emission devices for vacuum electronics applications [22–29].

Owing to the high surface-to-volume ratio, TMDs have excellent sensing performances. The $MoS_2$ sensitivity to NO and $NO_2$ has been demonstrated [30] with a detection limit of 0.8 ppm, but the exposure to other gases (studied from the theoretical point of view [31,32]) remains experimentally unexplored.

A considerable number of studies have been also devoted to the effects of the environment on mono- and few-layer TMD devices [33,34]. For instance, it has been shown that $WSe_2$ is very sensitive to pressure which can tune the conduction type [33]. Similarly, $PdSe_2$, which is a relatively new 2D material, has been demonstrated to be a good gas and pressure sensor [34].

A major difficulty that TMD-based nanoelectronics has to overcome is related to point defects as well as structural damages and dislocations, often generated during the fabrication, independently of the used process such as chemical vapor deposition (CVD) or mechanical exfoliation. Structural defects behave as charge traps and scattering centers, which modify the electronic properties of the devices and generate unwanted hysteresis and/or reduction of conductivity [3,35,36]. Hysteresis consists of a shift of the transistor transfer characteristic for consecutive forward/reverse gate voltage sweeps and changes the threshold voltage; it is an unwanted effect to circuits' designers, as it makes the transistor dependent on the biasing history. In spite of that, hysteresis can be conveniently exploited for the fabrication of memory devices [37–39], since it features two distinct and stable states, that can be used to define the bits of a memory cell. In this regard, it is interesting and important to understand the physical properties that control the hysteretic behavior of the transfer characteristic in TMD transistors.

In this paper, the electric properties of CVD-grown monolayer $MoS_2$ field effect transistors are studied under external stimuli, such as gate voltage, sweep delay time, pressure and pure gas environment, with particular attention to hysteresis. $MoS_2$ was selected among other TMDs because of his layer dependent bandgap over a wide range (1.2-1.9 eV), stability in air and mobility of few tens $cm^2V^{-1}s^{-1}$ when deposited on $SiO_2$ [4,7,8,40]. Chalcogen vacancies favor a natural n-type doping in $MoS_2$ and act as trap centers that enhance the hysteretic behavior in $MoS_2$ and others 2D TMDs [19,34,41]. We demonstrate exponential dependence of the hysteresis on the sweeping time and a linear dependence on the gate voltage range. We also show that exposure to gases such as oxygen, nitrogen and hydrogen at different pressure modifies the electrical properties of the devices [42], a feature that can be exploited for gas sensing purposes. In addition we prove that defective $MoS_2$ flakes are strongly sensitive to gases like methane ($CH_4$), as anticipated by DFT (density functional theory) studies [43]. We finally show that the wide hysteresis, especially if enhanced by gas adsorption, enables a two-bit memory device featuring a charge retention on the time scale of several minutes and an endurance on the order of hundreds of cycles.

**Methods:**

The $MoS_2$ flakes were grown in a three-zone split tube furnace (ThermConcept), purged with 500 Ncm³/min of Ar gas for 15 min to minimize the $O_2$ content. The growing $Si/SiO_2$ substrate was spin coated with a 1 % sodium cholate solution, then a saturated ammonium heptamolybdate (AHM) solution served as the molybdenum feedstock. The target material was placed in one zone of the three-zone tube furnace along with 50 mg of S powder, positioned upstream in a separate heating zone. The zone containing the S and AHM were heated to 150 °C and 750 °C, respectively. After 15 min of growth, the process was stopped, and the sample was cooled rapidly. Raman and atomic force microscopy (AFM) measurements were used to identify monolayer $MoS_2$ among the randomly distributed CVD-grown flakes. Using the substrate as common back-gate, we realized back-gated transistors by evaporating the drain and source electrodes on selected single layer flakes (see Fig. 1) by means of standard photolithography and lift-off process.

The contacts were made of Ti (10 nm) and Au (100 nm), deposited as adhesion and cover layers, respectively.

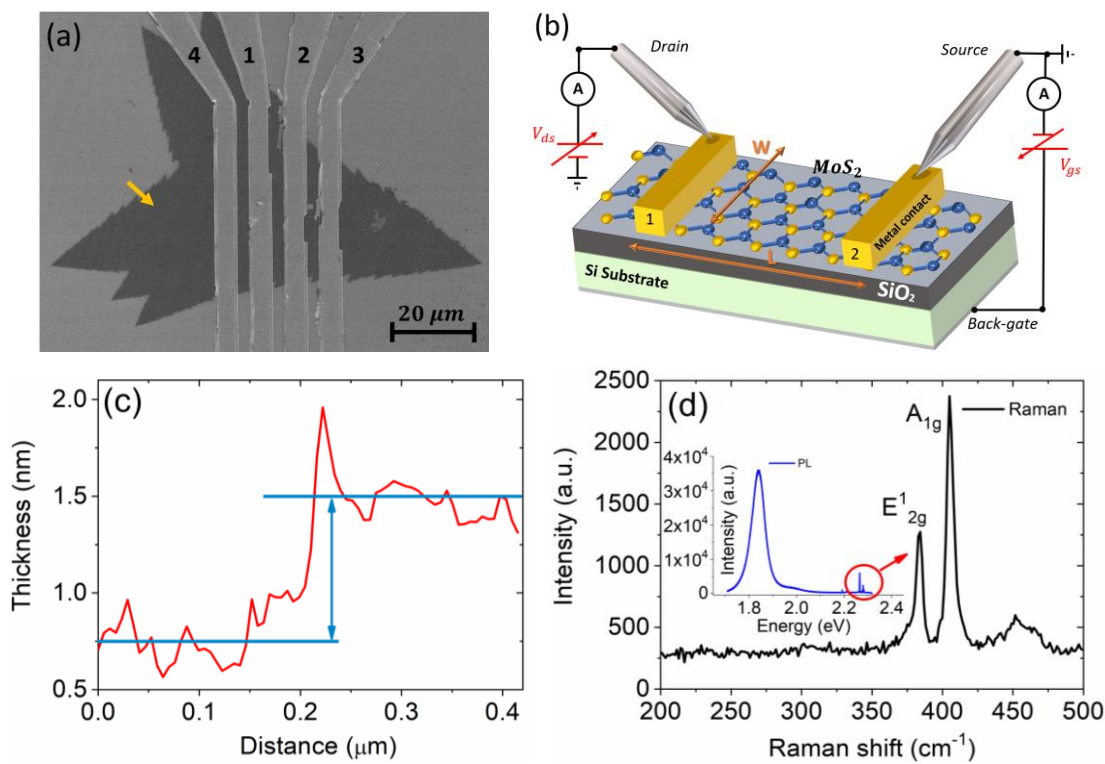

**Figure 1.** (a) SEM image of the selected flake with Ti/Au contacts; the yellow arrow indicates the AFM measurement direction. (b) Schematic of the device and the measurement setup; contacts 1 and 2 were chosen as the drain and source electrodes while scratched silicon substrate, covered with silver paint, operated as the back gate. The white arrows indicate the channel length (L ~ 5 µm) and width (W ~ 40 µm) of the device. (c) AFM measurement of the step height between the substrate and the flake. (d) Raman spectrum showing $E^1_{2g}$ and $A_{1g}$ modes separated by ∼ 20 cm$^{-1}$, typical of a monolayer $MoS_2$; the inset displays a photoluminescence peak at ∼ 1.85 eV energy bandgap, as expected for a monolayer.

In the following, the electrical characterization refers to the inner contacts 1 and 2 as displayed in Figure 1(a). Measurements were carried out inside a scanning electron microscope chamber (SEM, LEO 1530, Zeiss), equipped with two metallic tips, with curvature radius of about 100 nm (Figure 1(b)) and nanometric positioning capability, connected to a Keithley 4200 SCS (source measurement units, Tektronix Inc.), at room temperature and different chamber pressures, from ~ $10^{-6}$ Torr to 760 Torr.

The selected flake was characterized by atomic force microscopy and Raman/Photoluminescence (PL) measurements (Renishaw InVia Raman spectrometer with 532 nm laser wavelength). The results shown in Figure 1(c) and 1(d) prove that the flake is a single layer, as both the step height of ~ 0.75 nm and the large PL peak at ~ 1.85 eV, correspond to the typical height and bandgap values for a monolayer. We note that PL intensity and the AFM step height point to absence of intercalated water at the $SiO_2$ / $MoS_2$ interface, owing to high growth temperatures of 750 °C.

**Results:**

The electrical characterization of the device starts with the output and transfer characteristics, as shown in Figure 2, measured at $10^{-6}$ Torr. To avoid device damage, the channel current and the gate voltage were limited to 1 μA and 50 V, respectively. The output characteristics of Figure 2(a) show a slightly rectifying behavior typical of TMD transistors, where Schottky barriers between the channel and the contacts are easily formed [35,36,37]. Figure 2(b) confirms a normally-on n-type transistor with on/off ratio of ~ 6 orders of magnitude, featuring a subthreshold swing $SS = \frac{d(V_{gs})}{dLog(I_{ds})} \sim 1$ V/dec.

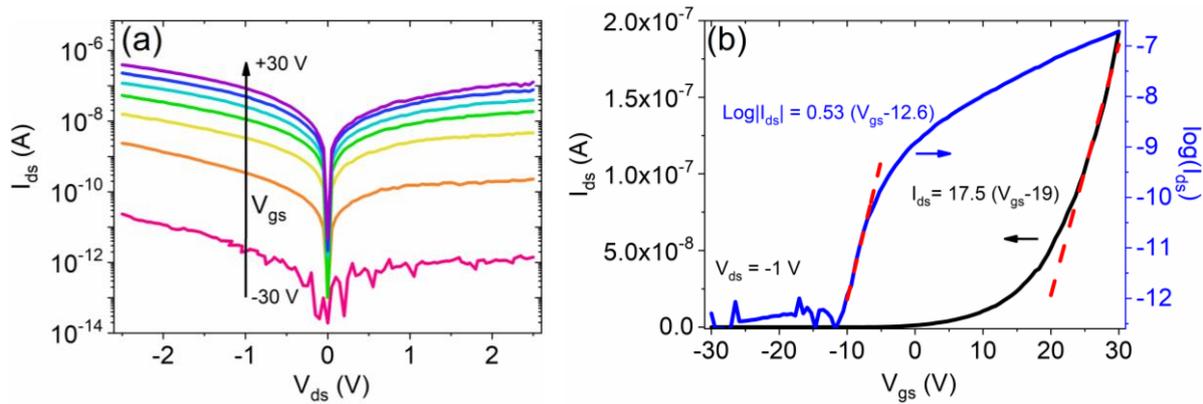

**Figure 2.** (a) Output and (b) transfer characteristics, measured at $10^{-6}$ Torr between contacts 1 and 2.

From the relation between mobility and gate voltage, $\mu = \frac{dI_{ds}}{dV_{gs}} \frac{L}{W\, C_{ox} V_{ds}}$, where $C_{ox}$ is the $SiO_2$ capacitance per unit area (11 nFcm$^{-2}$ for an oxide thickness of 300 nm), $V_{ds}$ the voltage bias and $dI_{ds}/dV_{gs}$ the transconductance (obtained as the slope of the transfer characteristic at high $V_{gs}$), we derived a mobility of ~ 1 cm$^2$V$^{-1}$s$^{-1}$.

This slightly low mobility, likely worsened also by the contact resistance [46], suggests the presence of trap states. Three types of trap states can be distinguished, as illustrated in Figure 3(a): The adsorbates on the $MoS_2$ surface (1), the intrinsic defects in the crystal structure of $MoS_2$ (2), and the extrinsic traps at the $MoS_2/SiO_2$ interface or into the $SiO_2$ dielectric layer (3). Each trap state is characterized by a trapping/detrapping time constant, which can be evaluated by several techniques.

A long annealing in high vacuum can remove most of the adsorbates (1) and allow the investigation of the effect of traps (2) and (3) only.

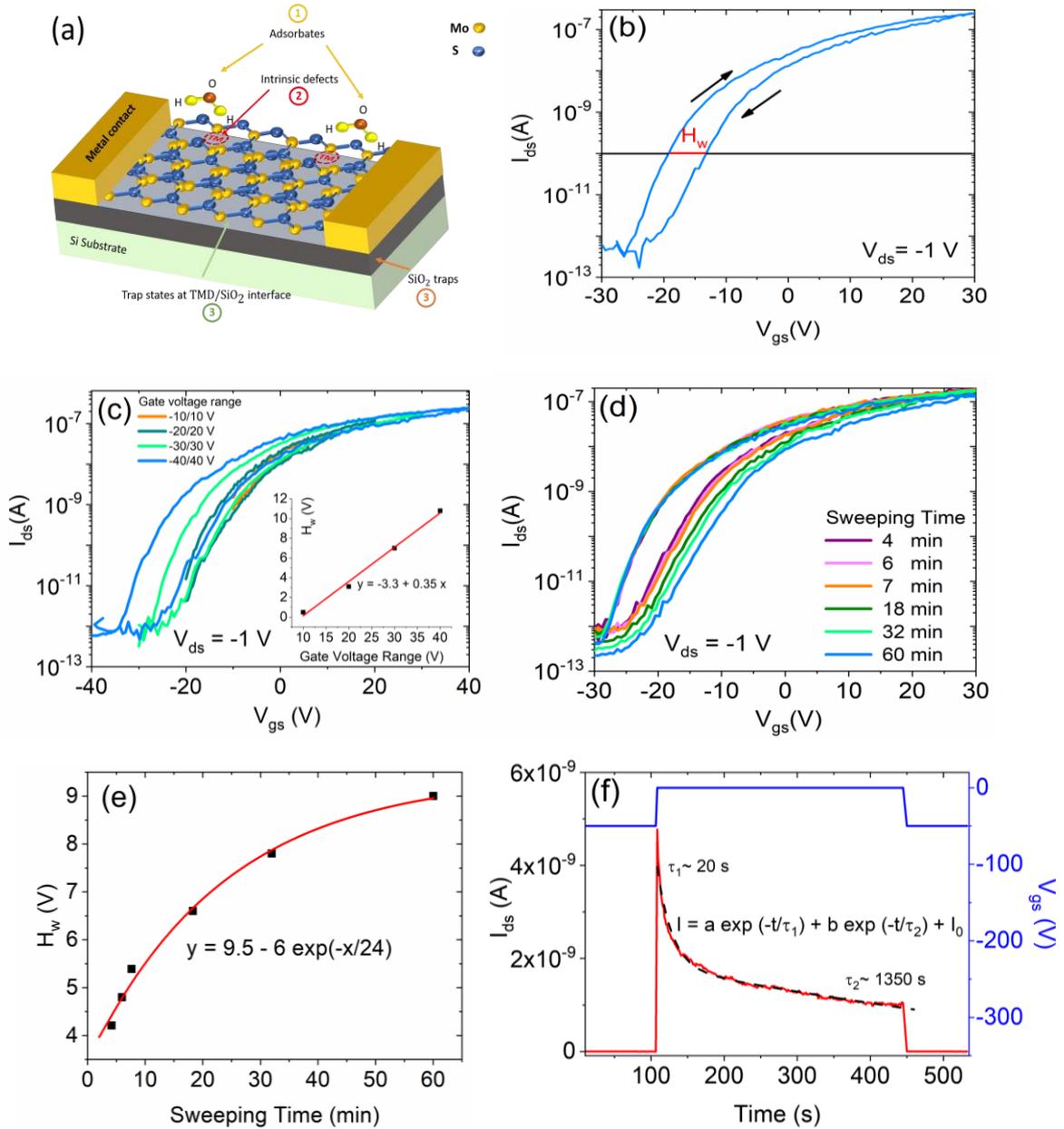

**Figure 3.** (a) Schematic of the different types of trap states present in a back-gate, unprotected MoS$_2$ transistor. (b) Transfer characteristic after 24 h anneal at 25 °C temperature and ~10$^{-6}$ Torr pressure. (c) Dependence of the transfer characteristic on the gate voltage range (the inset show the linear dependence of H$_W$), and (d) on sweeping time. (e) Exponential behavior of the hysteresis width as function of the sweeping time. (f) Experimental data (red curve) and fitting curve (black dashed) for the transient behavior of the transistor current during a gate pulse (blue curve) from -50 V to 0 V (at $V_{ds} = -1\ V$).

The transfer curve measured after 24 h at 25°C and ~10$^{-6}$ Torr shows a clockwise hysteretic behavior when the gate voltage is swept forth and back (Figure 3(b)). The right-shift of the transfer after a forward $V_{GS}$ sweep corresponds to negative charge trapping. Hysteresis can be characterized by a hysteresis width, H$_W$, defined as the difference of the gate voltages corresponding to the current of 0.1 nA. The hysteretic behavior is investigated as a function of the gate voltage range and the sweeping time in Figures 3(c) and 3(d). A linear dependence of H$_W$ on the gate voltage range is shown in the inset of Figure 3(c), while an exponential growth of H$_W$ with the sweeping time is reported in Figure 3(e). The linear increase of $H_W$ with the gate voltage range indicates that the trapped charge is proportional to gate potential, as expected considering that the trapping process load the capacitor formed by the $MoS_2$ channel and the Si substrate. The exponential growth of $H_W$ with the sweeping time [47] characterized by a long time constant of ~ 24 min, indicates a predominant role of slow (deep) traps either in $MoS_2$ or in the $SiO_2$ insulator (see (2) and (3) in Figure 3(a)).

More insight can be gained observing the transient behavior of the device under a gate voltage pulse. Figure 3(f) shows the channel current during a gate pulse of height 50 V and width ~400 s. The best fit is provided by a double exponential decay, $I_{ds} = a \exp\left(-\frac{t}{\tau_1}\right) + b \exp\left(-\frac{t}{\tau_2}\right)$, characterized by decay constants $\tau_1 \sim 20$ s and $\tau_2 \sim 1350$ s (i.e. $\sim 23\ min$), respectively. Such a behavior points to the presence of two types of electron trap states. The faster trapping is related to $MoS_2$ defects or $MoS_2/SiO_2$ interface trap states [48–50], while the slower trapping is ascribed to the filling of trap states inside the $SiO_2$ dielectric, controlled by the gate circuit. The exponential behavior of $H_W$ with the sweeping time indicates that the slower component is dominant when slow gate sweeps are applied. Following the procedure proposed by Xu et al. [35], we estimated a density of trap state in the $MoS_2$ structure of $\sim 10^{12}\ cm^{-2} eV^{-1}$ in agreement with other values reported in literature [51], using the transfer of Figure 3(b), corresponding to the fastest sweep (4 min).

We consider now the role of adsorbates by injecting several types of pure gases in the SEM chamber and investigating their effect on the electrical properties of the device.

We start monitoring the variation of transfer characteristic at different $O_2$ pressures (Figure 4(a)), as metrics to check its effect we chose the on-current and the channel mobility, as shown in Figure 4(b).

A monotonic change of the conduction parameters occurs while raising the pressure from $10^{-6}$ to 760 Torr. When the chamber is brought to atmospheric pressure in $O_2$ environment, a reduction of the on-current of one order of magnitude and an 80% decrease of the mobility are observed, owing to the adsorption of oxygen. Being highly electronegative, oxygen has an acceptor nature and traps electrons. This reduces the FET current and increases the coulomb scattering that degrades the channel mobility [36]. The compressive stress, applied by pressure to the $MoS_2$ channel, favors the interaction with the supporting $SiO_2$ dielectric and acts as a further mobility suppressor. The increased interaction with the substrate also facilitates charge transfer and contributes to hysteresis (see inset of Figure 4(a)).

Figure 4(c) shows the transfer characteristics measured in vacuum (~$10^{-6}$ Torr, solid pink curve) and in 760 Torr atmospheres of pure Ar, $H_2$, $O_2$, $N_2$ and $CH_4$, respectively. Such gases where chosen because of their great interest for gas detection and storage applications and for the availability of theoretical results based on DFT calculations. The hysteresis can be related to the adsorption energy ($E_{ads}$) of the various gases, defined as the difference between the total energy of the system ($E_{MoS_2+gas}$) and that of $MoS_2$ and the gas phase molecules alone ($E_{MoS_2}$ and $E_{gas}$ respectively)

$$E_{ads} = E_{MoS_2+gas} - E_{MoS_2} - E_{gas}$$

The adsorption energies were calculated through computational methods (DFT and ab initio studies), and the more negative is the adsorption energy the stronger is the interaction with $MoS_2$, that is the stability of the $MoS_2$+gas system [31,42,43,52–55].

The hysteresis width, evaluated at $10^{-11}$ A in Figure 4(c), as a function of the adsorption energy is displayed in Figure 4(d).

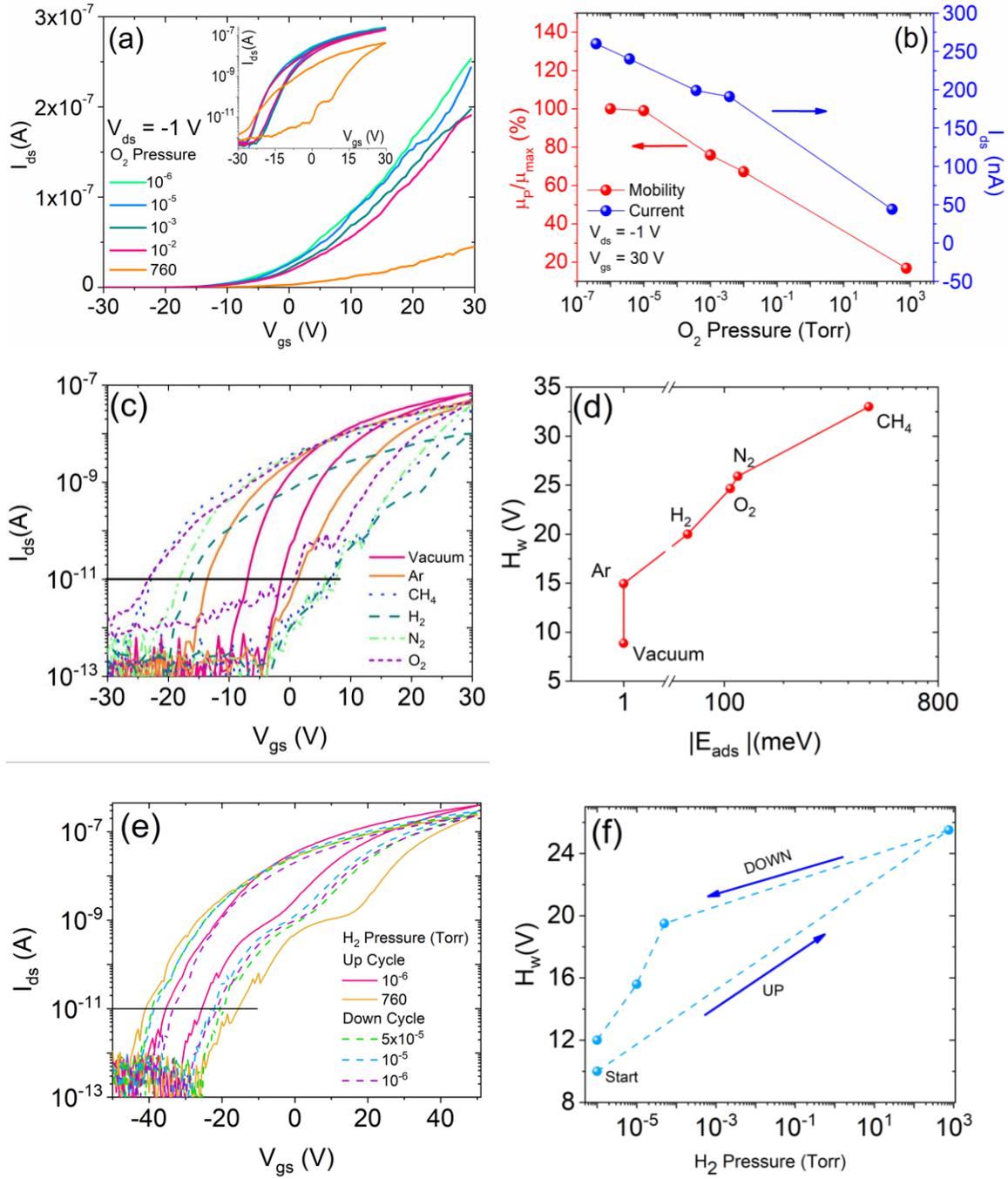

**Figure 4.** (a) Transfer characteristics at different $O_2$ pressures. (b) On-current (at $V_{gs} = 30$ V and $V_{ds} = -1$ V) and field effect mobility versus gas pressure. (c) Transfer characteristics measured under different gas atmospheres (Ar, $H_2$, $O_2$, $N_2$ and $CH_4$) at 760 Torr and in vacuum (<10$^{-6}$ Torr). (d) Hysteresis width versus gas adsorption energy on MoS$_2$ ($E_{ads}$ for $H_2$ and $O_2$ from ref. [31], N$_2$ from ref. [54], CH$_4$ from ref. [43]), as obtained from the sets of $I_{ds} - V_{gs}$ curves corresponding to $-30\,V \leq V_{gs} \leq 30\,V$ (Figure (c)). Given the linear relation between hysteresis and gate voltage range (displayed in Figure 3(c)), the $H_w - |E_{ads}|$ trend does not change for other gate voltage range choices. For display purposes, the adsorption energy corresponding to Ar and vacuum are set to 1 meV (no estimation of the E$_{ads}$ is currently available for Ar-MoS$_2$ system, while it results in the range 1-10 meV for Ar-carbon allotrope systems [56]). (e) Transfer characteristics and (f) hysteresis width when the pressure is raised from $10^{-6}$ Torr to 760 Torr and then decreased to $10^{-6}$ Torr, in $H_2$ environment.

The adsorption of Ar on $MoS_2$ has not been clearly investigated to date; for display purpose, we assumed $E_{ads} = 1$ meV in Figure 4(d). Ar could contribute to hysteresis, the width of which shows an increase of about 15 V with respect to vacuum, either by adsorption or by increasing the adhesion of the $MoS_2$ flake on the substrate, as pure pressure effect at 760 Torr [52], which increments the role of the interfacial and $SiO_2$ traps.

The comparison between transfer characteristics in different gas atmospheres (Figure 4 (c)) reveals that oxygen, nitrogen, hydrogen and methane strongly affect the $MoS_2$ electrical features, with a monotonic trend of $H_W$. This feature points to the suitability of $MoS_2$ FETs for gas sensing. Such an application is also corroborated by the observation that the effect of gases is reversible. Indeed, Figure 4(e) and 4(f) show that the transfer characteristic modified by the injection of a gas ($H_2$ in the example) returns about to the initial state when vacuum is restored. The effect of $H_2$ on $MoS_2$ nanostructures has been reported in connection with hydrogen storage application [42], while the effect of $CH_4$ has been anticipated by DFT studies dealing with low-defective few-layer $MoS_2$ crystals [43,54]. Indeed, the adsorption of $CH_4$ on the $MoS_2$ channel of the device under study confirms the presence of defects, since DTF study indicate positive adsorption energy (i.e. repulsion) of $CH_4$ on perfect monolayer $MoS_2$ [43]. Defects play an important role in the $H_W - |E_{ads}|$ relationship, as gas molecules are mainly adsorbed at the lattice defect sites, because of dangling bonds. The higher is the adsorption energy, the stronger is the interaction of the adsorbed molecules with $MoS_2$ lattice, which favors the charge exchange that is cause of hysteresis. Furthermore, dissociated and non-dissociated molecules act differently on $MoS_2$ electronic structure. Non-dissociated molecules, like $H_2$ and $N_2$, adsorbed on the $MoS_2$ surface, define additional trap centers without changing the band-structure. Conversely, dissociated molecules like $O_2$ modify the $MoS_2$ band structure by adding inter-band states ([57]). Moreover, the adsorption of $CH_4$ on $MoS_2$ surface generates a d-$MoS_2$/p-$CH_4$ orbital coupling, inducing a net transfer of charge (ref [43]).

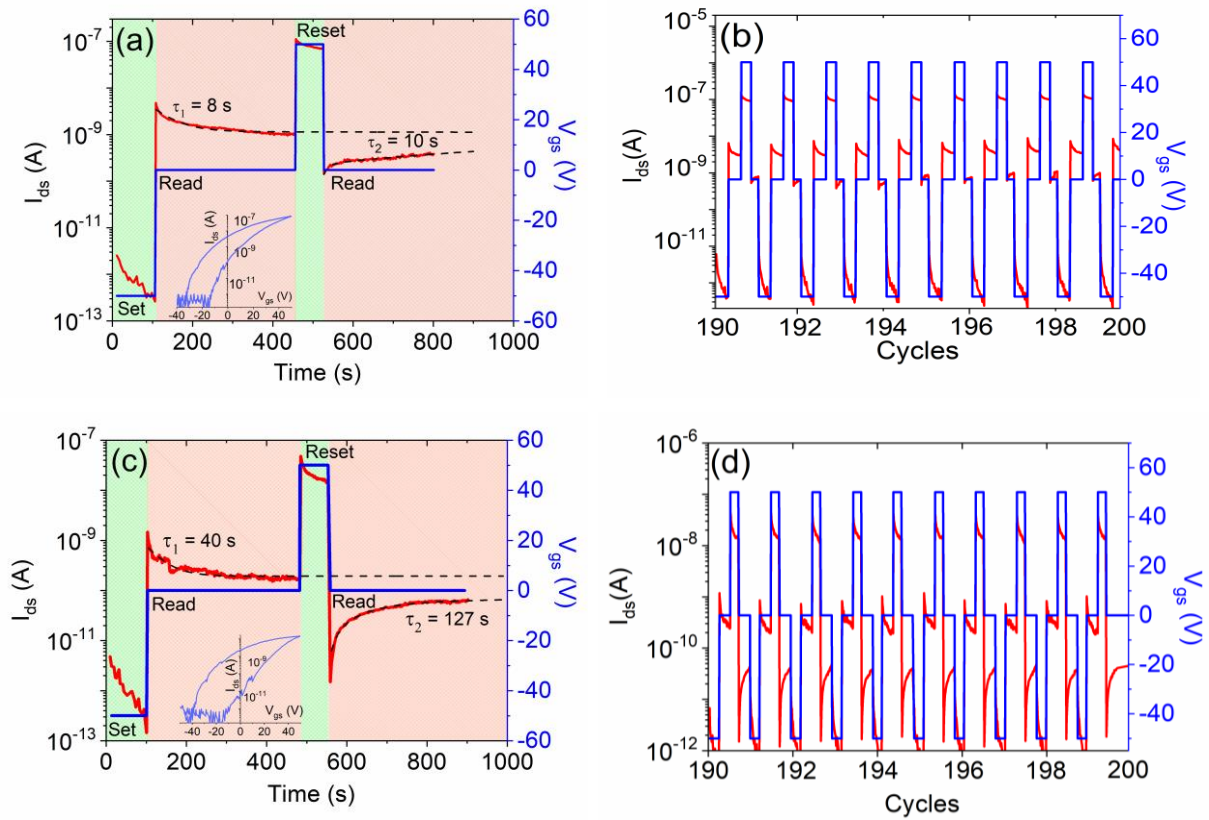

**Figure 5.** Single and multiple set-read-reset-read cycles for measurements performed (a-b) in vacuum ($\sim 10^{-6}$ Torr) and under $N_2$ atmosphere at a pressure of 760 Torr (c-d).

Finally, we show that the hysteresis in transfer characteristic can be exploited to realize a memory device. Figure 5 reports measurements for single and multiple set-read-reset-read cycles, both in vacuum ($\sim 10^{-6}$ Torr) and at atmospheric pressure. Figures 5(a) and 5(c) demonstrate that there is an order of magnitude current-level separation after ±50 V gate pulses and that the memory window (which is the separation between the two current levels) is kept constant after two hundred cycles (Figures 5(b) and 5(d)).

The device displays better performance under $N_2$ atmosphere, with more separated current levels. This suggests that annealing in selected gas environments can be a valid pretreatment in the fabrication of $MoS_2$ encapsulated devices, such as the ones covered by $Al_2O_3$[37], which are obviously more suitable for practical memory applications.

**Conclusions**

We have presented the electrical transport characterization of field-effect transistors with monolayer $MoS_2$ channel. The conductance shows an *n*-type behavior, with prevailing on-state over a wide voltage range and an intrinsic hysteretic behavior. Hysteresis has been investigated as a function of the range and the sweeping rate of the gate-voltage and has suggested that faster and slower components are involved, which are attributed to $MoS_2$ defects and $SiO_2$ traps, respectively. Most importantly, we have reported the effect of gas pressure and type on the transfer characteristics of the transistor. We have demonstrated that gas molecules can be adsorbed on defective $MoS_2$ surface, causing an increment of the hysteresis correlated with the adsorption energy of the system. Finally, we have confirmed the suitability of $MoS_2$ transistors as memory devices, especially when gases are stably adsorbed on the channel.

**Acknowledgments:** We acknowledge financial support by POR Campania FSE 2014–2020, Asse III Ob. Specifico l4, Avviso pubblico decreto dirigenziale n. 80 del 31/05/2016. We acknowledge support from the DFG by funding SCHL 384/20-1 (project number 406129719) and NU-TEGRAM (SCHL 384/16-1, projectnumber 279028710).

**Conflicts of Interest:** The authors declare no conflict of interest.